# On Understanding the Relation of Knowledge and Confidence to Requirements Quality

Razieh Dehghani[1][0000-0003-4619-0914], Krzysztof Wnuk[2][0000-0003-3567-9300],
Daniel Mendez[2,3][0000-0001-6300-o6635], Tony Gorschek[2][0000-0002-3646-235X],
Raman Ramsin[✉][1][0000-0003-1996-9906]

[1] Department of Computer Engineering, Sharif University of Technology, Tehran, Iran
[2] Department of Software Engineering, SERL Sweden/Blekinge Institute of Technology, Karlskrona, Sweden
[3] Fortiss GmbH, Germany
`rdehghani@ce.sharif.edu`, `Krzysztof.wnuk@bth.se`, `daniel.mendez@bth.se`, `tony.gorschek@bth.se`, `ramsin@sharif.edu`

**Abstract.** **[Context and Motivation]** Software requirements are affected by the knowledge and confidence of software engineers. Analyzing the interrelated impact of these factors is difficult because of the challenges of assessing knowledge and confidence. **[Question/Problem]** This research aims to draw attention to the need for considering the interrelated effects of confidence and knowledge on requirements quality, which has not been addressed by previous publications. **[Principal ideas/results]** For this purpose, the following steps have been taken: 1) requirements quality was defined based on the instructions provided by the ISO29148:2011 standard, 2) we selected the symptoms of low qualified requirements based on ISO29148:2011, 3) we analyzed five Software Requirements Specification (SRS) documents to find these symptoms, 3) people who have prepared the documents were categorized in four classes to specify the more/less knowledge and confidence they have regarding the symptoms, and 4) finally, the relation of lack of enough knowledge and confidence to symptoms of low quality was investigated. The results revealed that the simultaneous deficiency of confidence and knowledge has more negative effects in comparison with a deficiency of knowledge or confidence. **[Contribution]** In brief, this study has achieved these results: 1) the realization that a combined lack of knowledge and confidence has a larger effect on requirements quality than only one of the two factors, 2) the relation between low qualified requirements and requirements engineers' needs for knowledge and confidence, and 3) variety of requirements engineers' needs for knowledge based on their abilities to make discriminative and consistent decisions.

# 1 Introduction

Building software solutions requires achieving sufficient requirements quality. Requirements quality is affected by humans, processes and tools [1]. Requirements engineers' knowledge and their confidence are the two human-related factors that affect requirements quality. Researchers have previously assessed the effect of these factors separately. For example, it has been found that the effectiveness of interviews is affected by domain knowledge [2], [3]. Also, the relation between engineers' confidence and some specific types of requirements, such as safety, has been investigated [4].

Fig. 1 shows the research model used in this study. Hypotheses H4 and H5 refer to the effects of requirements engineers' knowledge and confidence on requirements quality. Since knowledge and confidence are interrelated [5], this research has focused on assessing the effects of knowledge deficit and a lack of confidence (hypotheses H6 and H7). Other relations, shown in Fig. 1, refer to the methods that have been used for assessing quality, knowledge, and confidence, as follows:

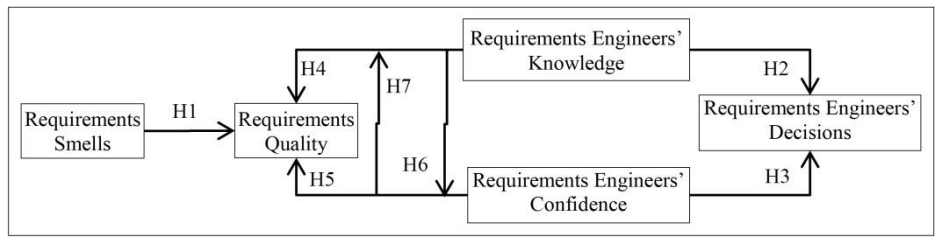

**Fig. 1.** Research Model

1) *Requirements Quality*: The ISO29148:2011 standard has provided a set of detailed principles for producing qualified SRS documents [6]. On this basis, Femmer et al. have defined the term *Requirements Smell* to assess quality [7]. Requirements smell is *"an indicator of a quality violation, which may lead to a defect, with a concrete location and a concrete detection mechanism"* [7]. Smells help find the location for low-qualified requirements. The location refers to the word/sentence, which violates the quality. For example, a vague adjective is a location for a low-qualified requirement because it might result in a misunderstanding about the requirement. It should be noted that the location might vary based on the product in which the requirements are stored. We have focused on SRS documents and used smells for assessing the quality of requirements (H1 in Fig. 1).

2) *Requirements Engineers' Knowledge*: This term is defined from a capability-based perspective. From this point of view, knowledge is "the potential to influence action" [8]. On this basis, *requirements engineers' knowledge* has the potential to influence the process of preparing SRS documents. Assessing the time that an individual spends in requirements engineering, namely experience, is a method for assessing knowledge. Besides, defects in decisions made by requirements engineers are symptoms of their level of expertise. In this research, low experience and inability to make discriminative and consistent decisions are considered as the symptoms of lack of enough knowledge [9], [10]. Discrimination and consistency have been defined from a comparative point of view [9]. On this basis, compared to novices, experts make more consistent and discriminative decisions throughout the requirements engineering process. Thus, H2 shows that inability in making discriminative and consistent decisions was used as the symptom for lack of enough knowledge.

3) *Requirements Engineers' Confidence*: This term refers to the feeling of trust about the SRS document that is prepared/reviewed/used. On this basis, uncertainty in making requirements engineering decisions was chosen as the symptom for low confidence (H3 in Fig. 1). This has been inspired by the results of Boness et al.'s research [11]. They have defined this term by proposing four criteria for refuting/warranting a claim about *requirements engineers' confidence* in goal-oriented requirements analysis. On this basis, we have proposed the following measures to assess confidence regarding various dimensions of requirements smells: depth of coverage, breadth of coverage, correctness, achievability, assumption, and accuracy. Analyzing the data about these criteria helps refute/warrant our claim about requirements engineers' certainty. It should be noted that uncertainty might occur regarding various features of requirements. We cannot claim that our research covers all these dimensions. However, by studying the research that has previously been conducted, we have tried to choose some specific dimensions of uncertainty regarding each dimension of requirements smells.

It should be noted that the methods we have used for assessing knowledge, confidence, and quality are context-independent [1], [9], [11]. However, some factors might affect the assessment. For example, the cultural features might affect requirements engineers' decisions [12].

The novelty of this research comes by addressing three issues: (1) in previous related work, requirements smells have not been traced so far to requirements engineers' knowledge and confidence, (2) abilities in making decisions have not been considered as symptoms of lack of requirements engineers' knowledge, and (3) interrelations between knowledge and confidence have not yet been considered.

Addressing these issues is important because: (1) requirements smells help trace the effect of low confidence and/or knowledge to a specific location(s) for low qualified requirements, (2) experience in requirements engineering, which refers to the time spent in academia and industry for requirements engineering, is not the only factor which affects individuals' knowledge; thus this research considers the skills in making requirements engineering decisions as well, and (3) ignoring the effect of low confidence or low knowledge yields wrong results and thus leads to inability to eliminate the causes for low quality.

The rest of this paper is organized as follows: next section provides an overview of the work related to this research; then, the method for conducting research and collecting data is explained; thereafter, results of analyzing the data are provided; and finally, the paper is ended by providing the conclusions and also suggesting some ways to further this research.

## 2     Background and Related Work

This work focuses on the intersection of three concepts: requirements quality, requirements engineers' knowledge and confidence. Femmer et al. have introduced requirements smells to assess the quality of SRS documents [7]. Similarly, Shanteau et al. and Boness et al. have respectively analyzed the individuals' knowledge and confidence by scrutinizing the decisions they make [9], [11]. Fig. 2 shows the terminologies and the relationships between the main concepts used in this work.

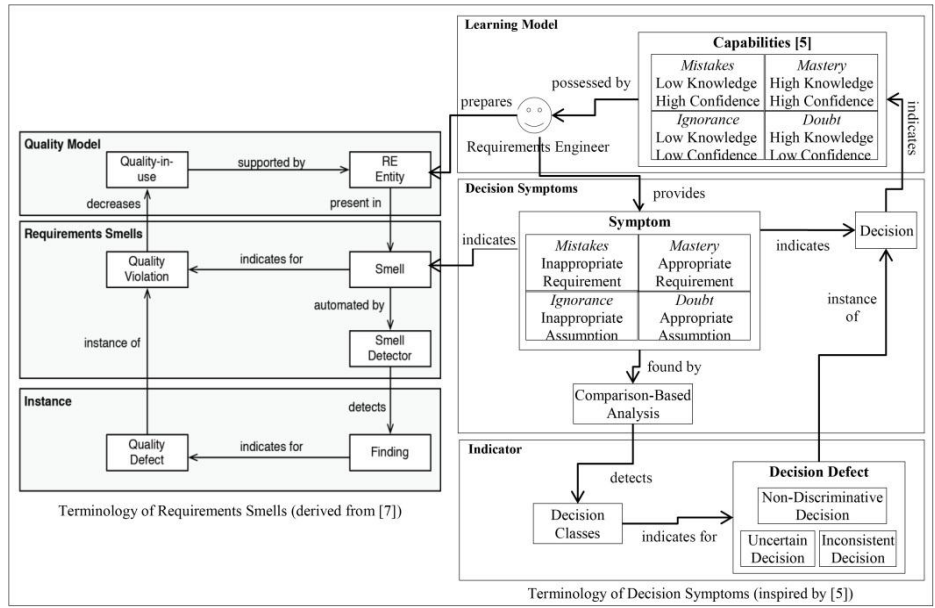

**Fig. 2.** Terminology of the Concepts Used in This Research

The left part of Fig. 2 is derived from Femmer et al.'s study [7] and presents the terminology for requirements smells. As explained in the first section, the requirements that do not follow the instructions provided by ISO29148:2011 standard [6], namely requirements smells, are low qualified [7]. We have categorized the research in the area of effects of smells as follows [13]:
- Effects of smells on *artifacts*: This category is concerned about the effects of smells on artifacts produced throughout the software development process. SRS is an example of an artifact affected by defects of natural languages [7].
- Effects of smells on *processes*: Research in this area addresses the effects of smells on development processes. As an example, the effect of requirements smells on test case design has been discussed in [14].

- Effects of smells on *people*: This area of research has been addressed indirectly. For example, Bjarnason et al. have provided a schema of requirements flow to depict the effect of requirements change on developers and customers [15].

Table 1 provides a high-level overview of various categories of smells. The source of the smells is provided in the third column. The smells might be related to requirements, the requirements process, or the time/place/logic/people-dependent conditions and constraints. It should be noted that the measures have been proposed based on the issue that is emphasized within the reference from which it has been elicited. More measures might also be elicited by other researchers.

**Table 1.** Categories of Requirements Smells

| Smell Dimension | Smell Category | Measure [reference] (ID-S#) |
|---|---|---|
| Requirement | Ambiguity | Probability of various interpretations regarding the meaning of requirement [7] (ID-S1) |
| Requirement | Incompleteness | Probability of having non-elicited requirements [16] (ID-S2) |
| Requirement | Inconsistency | Probability of having inconsistent requirements [16] (ID-S3) |
| Requirement | Redundancy | Probability of having redundant requirements [16] (ID-S4) |
| Requirement | Incorrectness | Probability of having semantically incorrect requirements [16] (ID-S5) |
| Requirement | Size | Probability of having compound requirements [16] (ID-S6) |
| Requirement | Size | Probability of having large SRS documents [16] (ID-S7) |
| RE Process | Analysis | Probability of having an inappropriate data collection method [16] (ID-S8) |
| RE Process | Analysis | Probability of having non-identified stakeholders [6] (ID-S9) |
| RE Process | Analysis | Probability of wrong judgment about criticality and risks [6] (ID-S10) |
| RE Process | Documentation | Probability of lack of explanation about "domain-specific and frequently occurring concepts" [16] (ID-S11) |
| RE Process | Documentation | Probability of having an incomplete glossary [16] (ID-S12) |
| RE Process | Verification | Probability of having inappropriate requirements verification method [16] (ID-S15) |
| RE Process | Validation | Probability of having requirements, non-traceable to stakeholders [16] (ID-S14) |
| RE Process | Validation | Probability of having non-defined "stakeholder requirements for validation" [16] (ID-S13) |
| RE Process | Management | Probability of having products non-traceable to requirements [6] (ID-S16) |
| RE Process | Management | Probability of having quality requirements without measures [6] (ID-S17) |

**Table 1 (Continued).** Categories of Requirements Smells

| Smell Dimension | Smell Category | Measure [reference] (ID-S#) |
|---|---|---|
| Time-Dependent Conditions and Constraints | Ambiguity | Probability of uncertainty about the order for satisfying the requirements [16] (ID-S18) |
| | | Probability of uncertainty about time for verification [6] (ID-S19) |
| | Incompleteness | Probability of having missing time-dependent conditions and constraints [16] (ID-S20) |
| Place-Dependent Conditions and Constraints | Ambiguity | Probability of having functionalities outside the boundaries of software architecture [6] (ID-S21) |
| | | Probability of making mistakes regarding system boundary [6] (ID-S22) |
| | | Probability of misalignment between stakeholder, system, and software requirements [6] (ID-S23) |
| | | Probability of having ambiguous "venue and environment for verification" [6] (ID-S24) |
| | Incompleteness | Probability of having unrecognized external elements (including regulations, culture, etc.) [6] (ID-S25) |
| | | Probability of having an incomplete configuration baseline [6] (ID-S26) |
| | | Probability of missing the constraints that affect the architecture [6] (ID-S27) |
| | Unavailability | Probability of inability in obtaining "items of information" [6] (ID-S28) |
| People-Dependent Conditions and Constraints | Ambiguity | Probability of uncertainty about stakeholders' preferences [16] (ID-S37) |
| | | Probability of uncertainty about interactions between users and systems [16] (ID-S38) |
| | Inconsistency | Probability of having wrong priorities regarding inconsistent stakeholders' requirements [16] (ID-S39) |
| | Incompleteness | Probability of specifying wrong individuals for conducting verification [16] (ID-S40) |
| | | Probability of having wrong supportive information about stakeholders [6] (ID-S41) |
| Logic-Dependent Conditions and Constraints | Ambiguity | Probability of unavailability of metadata regarding requirements [16] (ID-S29) |
| | | Probability of having open-ended sentences [16] (ID-S30) |
| | | Probability of having vague dependencies between requirements [16] (ID-S31) |
| | | Probability of having wrong overall integrity of requirements [6] (ID-S32) |
| | | Probability of having wrong estimations regarding goal satisfaction [16] (ID-S33) |

**Table 1 (Continued).** Categories of Requirements Smells

| Smell Dimension | Smell Category | Measure [reference] (ID-S#) |
|---|---|---|
| Logic-Dependent Conditions and Constraints | Ambiguity | Probability of having vague control flows [16] (ID-S34) |
| | | Probability of having vague logic behind optional requirements [16] (ID-S35) |
| | Incompleteness | Probability of having non-maintained rationale and assumptions [6] (ID-S36) |

The right part of Fig. 2 presents the terminology for decision symptoms. The "possessed by" arrow shows that each requirements engineer has some capabilities. Defects in making requirements engineering decisions are considered as symptoms of a lack of knowledge (including experience) or confidence. The defects can be classified as follows: 1) inappropriate assumptions are symptoms of ignorance because of low knowledge and confidence, 2) appropriate requirements indicate mastery in RE due to a high level of knowledge and confidence, 3) inappropriate requirements are symptoms of making mistakes because of low knowledge and high confidence, and 4) appropriate assumptions indicate doubt in RE due to high knowledge and low confidence.

It should be noted that the decisions might be inappropriate due to various reasons. That is why we have added a "decision classes" component in Fig. 2. As mentioned, we have selected three instances of defects, which are symptoms of inappropriateness, as follows:

1) *Uncertainty* is an indicator of the need for more confidence. Fig. 3 shows the model for assessing confidence. This is inspired by the procedures used in courts to refute/warrant a claim [17]. This method has previously been used for assessing confidence in requirements analysis, as well [11]. As shown, we first claim that the requirements engineer is not confident. Then, we look for the reasons through which we can warrant or refute our claim. To find the warranting and violating reasons, we have used the results of Boness et al.'s research (Table 2). As shown in Table 2, some measures have been proposed for assessing confidence regarding various dimensions of smells. It should be noted that these measures are not the complete set, and more measures might be added by researchers.

2) Inability to make *consistent* and *discriminative* decisions is the symptom of a lack of knowledge. It should be noted that experience is also a helpful factor for providing some assumptions about someone's knowledge, though it is not an accurate measure. We thus judged these assumptions by analyzing the decisions by using the CWS ratio [9], [10].

88

**Table 2.** Confidence Factors (Inspired by [11])

| Smell Dimension (Smell in) | Confidence Dimension | Measure (Confidence Factor) (ID-C#) |
|---|---|---|
| Requirement | What | Depth of Coverage: Confidence that the requirements have been adequately scrutinized in-depth (similar to refinement [11]) (ID-C1) |
| | | Breadth of Coverage: Confidence that the requirements have been adequately scrutinized in breadth (similar to engagement [11]) (ID-C2) |
| | | Correctness: Confidence that the requirements are correct (ID-C3) |
| | | Achievability [11]: Confidence that the requirements are achievable (ID-C4) |
| RE Process | How | Depth of Coverage: Confidence that the RE process has adequately covered the fine-grained RE tasks (ID-C5) |
| | | Breadth of Coverage: Confidence that the RE process has adequately covered the general RE process (ID-C6) |
| | | Correctness: Confidence that the RE process has been performed in the right way (ID-C7) |
| Time-Dependent Conditions and Constraints | When | Achievability: Confidence that the time-dependent conditions are achievable [11] (ID-C8) |
| | | Assumption: Confidence that the time-dependent constraints are sound [11] (ID-C9) |
| | | Accuracy: Confidence that the time-dependent conditions are specified (ID-C16) |
| Place-Dependent Conditions and Constraints | Where | Achievability: Confidence that the place-dependent conditions are achievable [11] (ID-C10) |
| | | Assumption: Confidence that the place-dependent constraints are sound [11] (ID-C11) |
| | | Accuracy: Confidence that the place-dependent conditions are specified (ID-17) |
| Logic-Dependent Conditions and Constraints | Why | Achievability: Confidence that the logic-dependent conditions are achievable [11] (ID-C12) |
| | | Assumption: Confidence that the logic-dependent constraints are sound [11] (ID-C13) |
| | | Accuracy: Confidence that the logic-dependent conditions are specified (ID-C18) |
| People-Dependent Conditions and Constraints | Who | Achievability: Confidence that the people-dependent conditions are achievable [11] (ID-C14) |
| | | Assumption: Confidence that the people-dependent constraints are sound [11] (ID-C15) |
| | | Accuracy: Confidence that the people-dependent conditions are specified (ID-C19) |

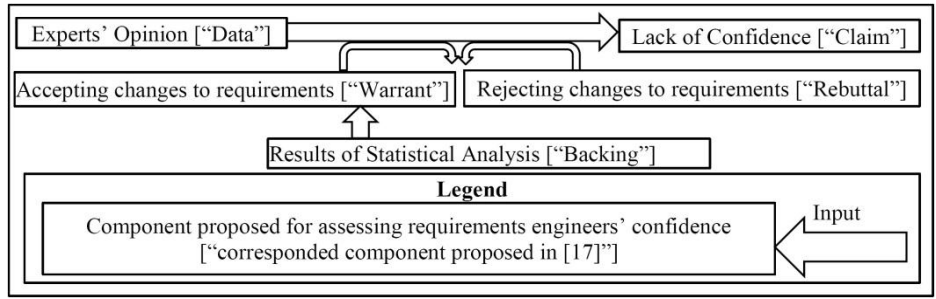

**Fig. 3.** Model for Confidence Assessment (Derived from [17])

## 3    Research Methodology and Data Collection

Fig. 4 presents the research steps followed in this work. First, we analyzed five SRS documents prepared by graduate students at the Blekinge Institute of Technology (BTH) in the course of their project work in Requirements Engineering and identified requirements smells in these documents. "The database should be reliable" is an example of a vague sentence (requirements smell). Next, we analyzed the project grading criteria to get aware of the requirements necessary for preparing the SRS documents and, thus, could not be considered as smells. Next, we designed the questionnaires to analyze students' knowledge and confidence, inspired by the Smith et al.'s four quadrants based on the level of human knowledge and confidence [5]: "Ignorance" (low knowledge and low confidence), "Doubt" (low confidence and high knowledge), "Mistakes" (high confidence and low knowledge), and "Mastery" (high confidence and high knowledge).

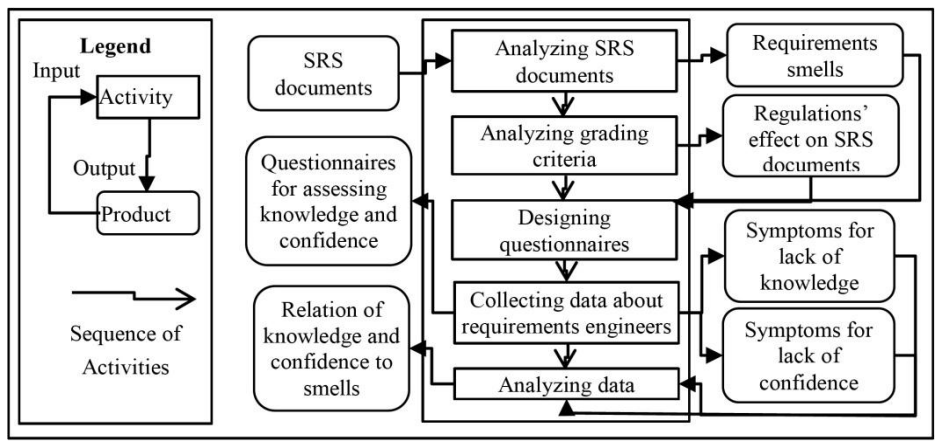

**Fig. 4.** Research Steps

The questions were answered by students who have prepared the SRS documents. Thus, the questionnaires encompass questions regarding certainty about specific smells found in the SRS documents, and students' abilities to make discriminative and consistent decisions. Examples of the questions are provided at the end of the paper, in the Appendix section. As an example, the students who have elicited the requirement about reliability *doubt* about the criteria by which reliability would be assessed. It should be noted that with the aim of alleviating the effect of environmental factors that might affect students' responses, the professors assured the students that the responses would not affect the grades.

Thus, as explained we have collected data in two steps:

1) *Analyzing SRS documents*: To find the smells, documents were analyzed by using the measures provided in Table 1. The results revealed a list of specific smells within each SRS document.
2) *Analyzing knowledge and confidence*: To assess students' confidence and knowledge regarding requirements, specific questions were designed for each group of students who have prepared the SRS documents. An instance of the instrument (questionnaire) we have designed is provided in the appendix section of this paper. After analyzing the data about students' knowledge and confidence, we could categorize the responses within the four mentioned quadrants. The method for analyzing knowledge and confidence is explained in the following paragraphs. First, we categorized the students based on their knowledge, and then we categorized their responses based on the response that shows the students' certainty/uncertainty.

To warrant or refute our claim about students' confidence, respondents were suggested to apply some changes to their documents, and they could "Agree" or "Disagree" with our suggestions. The changes were suggested in relation to the smells we have found in the first step. As shown in Fig. 3, agreeing with applying the changes was considered as a reason for warranting our claim about the lack of confidence. On the contrary, disagreeing with applying changes was a reason for rebutting our claim.

CWS ratio (Formula 1) was used [9], [10] for assessing the knowledge level. The abbreviation "CWS" comes from the names of individuals who have proposed it. This abbreviation "is used to establish that someone behaves more (high value) or less (low value) as an expert" [9]. In other words, this metric claims that judgments made by experts, in comparison with judgments made by novices, are more discriminating and consistent. According to [9], as shown in Table 3, for diagnostic decisions, there is a greater difference between decisions made by experts and novices, while for non-diagnostic ones, decisions are more similar.

$$\text{``CWS=Discrimination/Inconsistency'' [9]} \qquad (1)$$

**Table 3.** Difference between CWS Ratios (Derived from [9])

|  | **Important (Diagnostic)** | **Partially Important (Partially-Diagnostic)** | **Non-Important (Non-Diagnostic)** |
|---|---|---|---|
| Experts | A | B | C |
| Novices | D | E | F |
| | Difference between CWS ratio for experts and novices: A-D>B-E>C-F | | |

We have calculated the discrimination and inconsistency factors, provided in Formula 1, as follows: 1) first, we have investigated the number of years that respondents have experienced RE, and thus conducted a preliminary categorization regarding the respondents' expertise; 2) then, we have provided three categories of sentences, and respondents were asked to categorize them within the following classes: "Diagnostic" (important for eliciting the requirements), "Partially-diagnostic" (partially important for eliciting the requirements), and "Non-diagnostic" (not important for eliciting the requirements); 3) after that, we have measured the "Inconsistency" metric by calculating the "average of within-cell variances" ("low variance implies high consistency") [9]; 4) thereafter, the "Discrimination" metric was obtained by calculating mean square values ("High variance implies high discrimination") [9]; 5) after that, to calculate the CWS ratio (Formula 1), the discrimination metric was divided by the inconsistency metric; and 6) finally, we have reassessed our judgments about respondents' expertise by moving the students within categorizations so that we could make sure that experts are better in making consistent and discriminative decisions.

## 4    Results of Data Analysis

Eight groups of students (thirty-three individuals) participated in this study; however, we had to ignore the responses provided by three groups because more than half of the members of these groups did not fill in the questionnaires. The following paragraphs respectively discuss the results of analyzing the data collected for finding the relation of smells to confidence, knowledge, and both knowledge and confidence.

1) *Analyzing data about confidence*: Table 4 provides an example of the responses we have received to assess confidence; rows represent question numbers, and columns represent respondent numbers. As shown, the changes suggested for four questions were agreed upon by at least half of the group members. Table 5 shows the number of respondents who agreed with making the changes suggested for each group; rows represent question numbers, and columns represent group numbers. As shown, we found that except for eight changes, all other ones were agreed to be applied by at least half of the respondents. Thus, it is concluded that the students have confirmed that they are not confident regarding the requirements smells we have found.

**Table 4.** Example of Responses to Questions for Assessing Confidence (Group 1)

|     | R1  | R2  | R3  | R4  | TNA |
|-----|-----|-----|-----|-----|-----|
| Q1  | "0" | "0" | "1" | "1" | 2   |
| Q2  | "0" | "0" | "0" | "0" | 0   |
| Q3  | "1" | "1" | "1" | "1" | 4   |
| Q4  | "1" | "1" | "1" | "1" | 4   |
| Q5  | "1" | "1" | "0" | "0" | 2   |

**Legend:**
"1" refers to agreeing with applying the change, and "0" refers to disagreeing with applying the change
TNA stands for Total Number of Agreements

**Table 5.** Number of Respondents Who Agreed with Making the Changes (Groups 1-5)

|      | G1   | G2   | G3  | G4  | G5  |
|------|------|------|-----|-----|-----|
| Q1   | 2    | 5    | 6   | 5   | 6   |
| Q2   | **0**| 4    | 5   | 3   | 3   |
| Q3   | 4    | **1**| 4   | 4   | 5   |
| Q4   | 4    | 4    | 3   | 5   | 5   |
| Q5   | 2    | 4    | 5   | 5   | 4   |
| Q6   | 3    | 5    | 4   | 4   | 3   |
| Q7   | **1**| 5    | 4   | 3   | 4   |
| Q8   | 4    | 5    | 6   | 5   | 6   |
| Q9   | 3    | 5    | 6   | 4   | 5   |
| Q10  | 3    | 3    | 5   | 5   | 5   |
| Q11  | **1**| 3    | 3   | 3   | 4   |
| Q12  | 2    | **2**| 5   | 3   | 4   |
| Q13  | 4    | **1**| 6   | 3   | 4   |
| Q14  | **1**| 3    | 3   | 3   | **2**|

**Legend:** Bold underlined numbers indicate that less than half of the respondents agreed with making the change.

2) *Analyzing data about knowledge*: Table 6 provides the results of calculating the CWS ratio. It should be noted that we have calculated this metric by using three pre-classification methods as follows: 1) timespan of experience in academy environments, 2) timespan of experience in non-academy environments, and 3) the total timespan of experience in both academic and non-academic environments. What we found was that for the third type of pre-classification, in comparison with the other two pre-classification methods, the decisions made by experts and novices are more clearly discriminated (as specified in Table 3).

**Table 6.** CWS Ratio (Pre-classification was made based on the total timespan of experience in both academic and non-academic environments)

| Category | Important (Diagnostic) | Partially Important (Partially-Diagnostic) | Non-Important (Non-Diagnostic) |
|----------|------------------------|--------------------------------------------|--------------------------------|
| **Group 1** | | | |
| **Experts** | 4 | 0.25 | 0 |
| **Novices** | 0.8 | 0.13 | 0 |
| **Result:** 4-0.8>0.25-0.13>0-0 | | | |
| **Group 2** | | | |
| **Experts** | 5 | 0.34 | 0 |
| **Novices** | 0.7 | 0.21 | 0 |
| **Result:** 5-0.7>0.34-0.21>0-0 | | | |
| **Group 3** | | | |
| **Experts** | 2.5 | 0.45 | 0.1 |
| **Novices** | 0.4 | 0.10 | 0 |
| **Result:** 2.5-0.4>0.45-0.10>0.1-0 | | | |

13**Table 6 (Continued).** CWS Ratio

| Category | Important | Partially Important | Non-Important |
|---|---|---|---|
| **Group 4** | | | |
| **Experts** | 3 | 0.24 | 0.03 |
| **Novices** | 0.7 | 0.10 | 0 |
| **Result:** 3-0.7>0.24-0.10>0.03-0 | | | |
| **Group 5** | | | |
| **Experts** | 4.3 | 0.38 | 0 |
| **Novices** | 0.6 | 0.19 | 0 |
| **Result:** 4.3-0.6>0.38-0.19>0-0 | | | |

3) *Analyzing data about both knowledge and confidence*: In total, 33 respondents have specified their opinion regarding 14 smells. Thus, we have received 462 (33 multiplied by 14) responses regarding smells. Each of these responses falls into one of the knowledge-confidence quadrants, based on the evaluation made regarding the respondents. As shown in Fig. 5, the "Ignorance" quadrant encompasses the most responses, which means that a combined lack of knowledge and confidence has the most negative effect on requirements smells, and thus requirements quality.

| Mistakes<br>(Low Knowledge<br>High Confidence)<br><br>155 responses | Mastery<br>(High Knowledge<br>High Confidence)<br><br>43 responses |
|---|---|
| Ignorance<br>(Low Knowledge<br>Low Confidence)<br><br>172 responses | Doubt<br>(High Knowledge<br>Low Confidence)<br><br>92 responses |

**Fig. 5.** Number of Responses in Knowledge-Confidence Quadrants

Looking at the results, we draw attention to the following issues:
1) Uncertainty about requirements is an indicator of low-qualified requirements. Practitioners can check requirements engineers' confidence to find requirements that are potentially low-qualified. Besides, finding and classifying the reasons for uncertainty is an area of research which is needed to be addressed by researchers.
2) Various requirements engineers might elicit different requirements for a unique software system. This is due to the difference in their knowledge. The CWS ratio helps find the differences. Project managers can use this metric to categorize the employee and plan to improve their skills in RE. Besides, researchers should identify the determinative decisions which should be consistent and discriminative.
3) Requirements quality is affected by a collection of factors. Not only the factors but also their relations affect quality. Due to the interrelation between knowledge and confidence, a lack of confidence simultaneously with a lack of knowledge increases the probability of low quality. Researchers should explore such relations, and practitioners should beware of the simultaneous effects of interrelated factors.

## 5 Conclusions and Future Work

In this research, we explored the relationship between knowledge and confidence, and requirements quality. For this purpose, we have first analyzed five SRS documents developed by the students of Blekinge Institute of Technology. The analysis aimed to find the low qualified requirements, which was done using a set of criteria named requirements smells. In the next step, students' knowledge and confidence were assessed by analyzing their abilities to make discriminative, consistent, and certain decisions. Finally, we have classified smells based on individuals' knowledge and confidence. We found that most smells fall into the class with a lack of confidence and knowledge.

Thus, requirements for smells might be considered as symptoms of a lack of knowledge and/or confidence. Project managers can use this information (the relation between requirements smells, knowledge, and confidence) to find the areas in which some training mechanisms should be used to improve requirements engineers' skills. As an example, they might decide to hold some workshops to improve requirements engineers' skills. The training materials used within these workshops can be decided on this basis.

This research is novel mainly due to considering the interrelation between knowledge and confidence, using a decision-based comparative method for analyzing knowledge, and analyzing requirements quality based on specific symptoms for low quality. However, conducting one experiment in one academic environment is not enough for approving the relations, and more cases should be investigated to approve the results in general. We aim to further this research by conducting more experiments through which we can collect more data.

## Appendix

Some examples of the questions that we have designed are provided herein. The questions in the following three sections are respectively aimed at assessing confidence, analyzing domain knowledge, and investigating knowledge in RE.

*Section A:* Imagine that a company manager has studied the SRS document that you have prepared for this project, and you are invited to join a team to help develop the system for which you have elicited the requirements. For the first step, the manager provides the following claims about your document and asks you to address them. Please indicate if you agree/disagree?

1) Regarding the following requirement, much more detail is required and still, it should be refined. DL1: "The information shall be presented using HTML5.2 and CSS3 languages." Agree ☐   Disagree ☐
2) More detail about time-dependent conditions and constraints are required for these requirements: PR5: "The web application shall offer the functionality

   of registration in the web-app.", and PR4: "The web application shall offer the functionality of login in the web-app." Agree ☐    Disagree ☐
3) You are not sure about the appropriate time for verifying the requirements. Agree ☐    Disagree ☐
4) Policy and regulations have been provided. The effects of cultural elements should also be discussed. Agree ☐    Disagree ☐
5) You are not sure about the dependency between some requirements. For example, it seems that some issues regarding the dependency between the following requirements are not explained: PR1:"The web application shall offer the functionality of adding a new movie review.", and PR2:"The web application shall offer the functionality of rating a movie." Agree ☐    Disagree ☐

*Section B:* Please answer the following questions:
- How many *industrial* (*non- academic*) projects have you been engaged in to develop a software system, the same as the system you have engineered requirements for (in the role of a project manager, programmer, etc.)?
- How many *academic* projects have you been engaged in to develop a software system, the same as the system you have engineered requirements for (in the role of a project manager, programmer, etc.)?

Please categorize the following issues as important, partially-important, and non-important in selecting the most suitable requirements prioritization techniques.
- Type of requirement (functional/non-functional)
- Support for evaluating requirements
- Caring about requirements dependencies
- Support for coordinating various stakeholders' requirements
- The number of requirements that should be prioritized

*Section C:* Please answer the following questions:
- How many *industrial* (*non- academic*) projects have you been engaged in for eliciting requirements?
- How many *academic* projects have you been engaged in for eliciting requirements?

Please categorize the following issues as important, partially-important, and non-important for selecting the most suitable requirements elicitation techniques.
- Complementary requirements elicitation techniques that are required to be applied.
- Number of requirements that would be elicited by the technique(s) chosen.
- People-dependent factors (such as culture).
- The time that it would take to elicit the requirements.

## Acknowledgement

We would like to acknowledge that this work was supported by the KKS foundation through the S.E.R.T. Research Profile project at Blekinge Institute of Technology and the SERL Lab.